# "La Ciencia en los Cuentos": Análisis de las imágenes de científico en literatura juvenil de ficción

Alejandro Pujalte[1], Alejandro Gangui[1,2,3] y Agustín Adúriz-Bravo[1]

[1]CeFIEC-Instituto de Investigaciones Centro de Formación e Investigación en Enseñanza de las Ciencias, Facultad de Ciencias Exactas y Naturales, Universidad de Buenos Aires, Argentina.

[2]IAFE-Instituto de Astronomía y Física del Espacio, Buenos Aires, Argentina.

[3]CONICET-Consejo Nacional de Investigaciones Científicas y Técnicas, Argentina.

E-mail (primer autor): ap_pujalte@yahoo.com.ar

**Resumen:** Dentro del marco general de las investigaciones acerca de las "imágenes de científico" sostenidas en diferentes grupos de personas, ámbitos y manifestaciones culturales, pueden ser relevantes los estudios referidos al carácter de tales imágenes en la literatura de ficción. Nuestro foco está puesto en una serie de cuentos producidos por jóvenes estudiantes de los últimos años de la escuela secundaria (16 a 18 años), con la intención de vincular algunos rasgos estereotípicos del científico especialmente frecuentes en esos escritos con los que subyacen a otras producciones culturales, en tanto que estas últimas se pueden constituir como fuentes probables de las imágenes utilizadas. Sostenemos que muchas de las representaciones acerca del científico son emergentes de una imagen de ciencia inadecuada desde el punto de vista de la enseñanza de las ciencias que se constituye como un auténtico *obstáculo*, pues se corresponde con un desinterés por las asignaturas científicas. Esto último, sumado a otros condicionantes multicausales, contribuiría a un estancamiento de la matrícula en las carreras de ciencia y tecnología en relación con otros recorridos académicos, en Iberoamérica en general y en especial en la Argentina.

**Palabras clave:** imagen de científico, imagen de ciencia, literatura juvenil, ficción científica, obstáculos, educación científica de calidad.

# 'Science in Tales': Analysis of the images of scientist in young people's fictional literature

**Abstract:** In the framework of the research on the 'images of scientists' that appear within different groups, contexts and cultural manifestations, studies related to such images in fictional literature can be relevant. We focus on a set of tales produced by young students of the last years of secondary school (aged 16-18). Our aim is to connect some traits of the stereotypical scientist that appear in those texts with the traits underlying other cultural productions, since those productions may constitute the probable source of the stereotype. Our claim is that many common representations of the scientist emerge from an image of science that is inadequate from the point of view of science teaching. Such image constitutes a genuine *obstacle*, since it correlates to a lack of interest towards science courses. This lack of interest, added to other multi-causal conditionings, would contribute to stagnation in the number of students pursuing science and technology careers as compared to other academic paths, in Ibero-America in general and especially in Argentina.

**Introducción**

Son muy conocidos los abordajes analíticos realizados sobre las imágenes de ciencia y de científico presentes en las creaciones literarias (novelas, cuentos, obras de teatro...), sobre todo aquellas de los siglos XVIII, XIX y XX. En nuestro caso, el material sobre el que detuvimos nuestra mirada está integrado por una serie de cuentos escritos por jóvenes estudiantes de últimos años de escuela media (16-18 años), producto de un concurso denominado "La Ciencia en los Cuentos". Más allá de la estimación de la calidad literaria de las obras integrantes de esa producción (de la que ya dieron cuenta los respectivos jurados en su momento), tales obras nos interesaron como insumo para el análisis desde la perspectiva de la llamada *naturaleza de la ciencia* (NOS).

En nuestro grupo de investigación, una de las líneas de acción en los últimos años es la referente a las indagaciones de las imágenes de ciencia y de científico en diferentes audiencias, en la que se inscribe el proyecto de tesis doctoral del primer autor de este trabajo (A.P.). Está fuera de discusión en la comunidad de didactas de las ciencias naturales la importancia de relevar las imágenes de ciencia y de científico que los actores educativos traen consigo. En ese sentido, abundantes trabajos han sido realizados buscando indagar las visiones acerca de la ciencia que sostienen, por un lado, el profesorado de ciencias en formación y en activo, y por otro, los/as estudiantes (presentamos una reseña de esos trabajos más adelante en este mismo artículo). Desde hace más de cincuenta años se viene indagando las imágenes de científico que traen el estudiantado y el público general, imágenes que podemos considerar un emergente de la imagen *folk* de ciencia socialmente instalada. Las formas de indagar han sido muy variadas: cuestionarios abiertos y cerrados, entrevistas, dibujos figurativos y metafóricos, etc.

Consideramos que los cuentos producidos por los jóvenes tomados como población de nuestro estudio pueden constituirse en un nuevo corpus con características interesantes. Las 147 historias sobre las que hemos trabajado surgieron de un concurso que apuntó a motivar a los/as jóvenes para que, investigando algún aspecto de la ciencia y usando su imaginación, elaboraran una narrativa de ficción. Las imágenes de ciencia que subyacen en los cuentos presentados al concurso están llenas de matices que permiten lecturas complejas sobre la ciencia que se enseña y que se aprende.

**¿Qué es "La Ciencia en los Cuentos"?**

Se trata de un concurso literario realizado en la Argentina, con versiones anuales desde el año 2006 al 2011, dirigido a jóvenes estudiantes de escuela media. Es el propósito de este certamen[1]:

> (…) motivar a los jóvenes para que investiguen algún aspecto de la ciencia que los fascine, para que desarrollen una idea, usen su imaginación, y expresen el resultado de sus meditaciones con palabras cuidadas en una obra que sea a la vez rigurosa como documento científico y literariamente atractiva.

Como consigna general, allí mismo se dice que:

> No se trata de un artículo de divulgación científica. No se trata tampoco solo de una narración de ciencia ficción. Se trata de usar la imaginación, de ser claro y conciso; se trata de disfrutar de la escritura, de mostrar dominio de la ciencia que se describe, y de ser original.

---

[1] Según las Bases del Concurso "La Ciencia en los Cuentos", disponibles en: http://cms.iafe.uba.ar/gangui/difusion/concurso/

Hemos tomado para nuestro análisis 147 relatos surgidos de este certamen con los siguientes propósitos:

1. Indagar cuál es la imagen de científico que revelan estas narraciones. A través de ella, hacer hipótesis acerca de la imagen de ciencia del autor o la autora.

2. Establecer los posibles nexos con las imágenes de científico presentes en otras producciones culturales.

3. Analizar críticamente, desde la perspectiva NOS, qué relaciones hay entre la imagen de ciencia que subyace en los cuentos y la ciencia que se enseña y que se aprende en las aulas.

**Las indagaciones de la imagen de ciencia y científico: Algunos antecedentes**

Desde hace muchos años se viene tratando de indagar las imágenes que los jóvenes tienen respecto de la ciencia. Uno de los trabajos pioneros en este sentido es el de la antropóloga Margaret Mead (Mead y Metraux, 1957), quien mediante cuestionarios y entrevistas buscó elicitar la imagen de ciencia que sostienen estudiantes de secundaria, en un estudio que involucró un número muy importante de jóvenes (cerca de cuarenta mil) de distintas escuelas de Estados Unidos. Básicamente, los resultados mostraron un reconocimiento del valor de la actividad científica y de los productos de la ciencia en beneficio de la humanidad. Ahora bien, cuando de alguna manera se ponía a los sujetos en función de verse involucrados con la ciencia como elección personal, en tanto trayecto formativo o futura profesión, la desidentificación y el rechazo como opción de vocación posible resultaron muy evidentes.

A continuación (figuras 1 y 2) se presentan dos esquemas que reseñan los rasgos estereotípicos que relevara Mead en su investigación.

--- Figura 1 más o menos aquí ---

--- Figura 2 más o menos aquí ---

Ha transcurrido bastante tiempo desde esa investigación y no parece que esta imagen haya cambiado significativamente. El estereotipo del científico sigue incólume y no constituye un modelo al cual la mayoría del estudiantado quiera adherir. Desde este estudio en adelante, las posteriores indagaciones realizadas por otros investigadores abonaron estas conclusiones

(Beardslee y O'Dowd, 1961; Brush, 1979); así, "[m]ientras que la mayoría de la gente expresa respeto y admiración por los científicos, el estereotipo dominante desalienta a quienes no se identifican con él cuando se piensan a sí mismos como científicos" (Leslie-Pelecky et al., 2005).

Muchos estudios han revelado la temprana formación de esta imagen; la evidencia muestra que ya a los seis o siete años las niñas y niños producen estas representaciones (Newton y Newton, 1988). De hecho, son ampliamente conocidos los trabajos que resultan de la aplicación del DAST (*Draw a Scientist Test*: Chambers, 1983) a niños de los primeros años de la educación primaria; en ese test se les pide que dibujen a una persona que se dedica a la ciencia en su ambiente de trabajo. Una de las cuestiones más abordadas en estos análisis es la percepción de género en los dibujos, notándose una amplia recurrencia al científico varón. Sin embargo, existen matices: se ha señalado que a edades tempranas suele haber una representación del propio género, y, por una cuestión madurativa, las niñas suelen hacer dibujos más fácilmente reconocibles como pertenecientes al género femenino que en el caso de los niños, donde a veces esa distinción no se hace sencilla. Los varones, en general, tienden a encasillar más al científico dentro del estereotipo (Losh et al., 2008).

Cabe destacar los proyectos multinacionales abocados a indagar qué es lo que sucede con estas percepciones y actitudes del estudiantado en diferentes países; uno de tales proyectos ha sido "The SAS Study", dirigido por el Svein Sjøberg, de la Universidad de Oslo (Sjøberg, 2000). En ese estudio, las niñas y niños de los países en vías de desarrollo presentaron una visión sobre la ciencia y la tecnología mucho más positiva que la de aquellos de los países ricos. Mientras que en estos últimos países, los niños (principalmente varones) representaban al científico como una persona cruel y loca, en los países en vías de desarrollo los consideraban ídolos y héroes. Se señala que en los países desarrollados existe un problema con la imagen pública de la ciencia y una creciente preocupación por la baja de la matrícula en las carreras relacionadas con la ciencia y la tecnología. Al preguntarse acerca de esta suerte de contradicción –dado que estos niños, adolescentes y jóvenes viven en sociedades basadas en el conocimiento científico y tecnológico–, los investigadores sostienen una explicación posible: que se trata del resultado de una "baja" comprensión pública de la ciencia, atribuible a una enseñanza de mala calidad, como también de una mala imagen reflejada en los medios de comunicación masivos.

Respecto del DAST y su aplicación, ha habido reformulaciones, algunas de ellas a partir de ciertas críticas. Tal es el caso del trabajo llevado adelante por Manzoli y colegas (2006), donde se señalan algunas consideraciones respecto de este test en los siguientes sentidos: 1. que no revela cómo y de dónde se construye el estereotipo, 2. que da una imagen estática que no permite al estudiante dar detalles de la ciencia como proceso, y 3. que no queda del todo claro si es que el DAST mide los estereotipos del estudiantado o bien sucede que los científicos son dibujados así para hacerlos reconocibles como tales por quienes demandan el dibujo. El estudio de referencia se realizó en Italia, con niñas y niños de 8 y 9 años; en él, además de pedírseles el dibujo se les solicitó que inventaran una historia que tuviera al científico como uno de sus protagonistas. Otros investigadores en el tema también han echado mano de este recurso, como en el caso del trabajo de Reis y colegas (2006). A partir de los resultados, los autores concluyen, por un lado, que las representaciones que niñas y niños manifiestan van más allá del estereotipo, que en realidad es usado sólo como "esqueleto" sobre el cual volcar aspectos que tienen más que ver con lo que captan de los medios, y por otro, que esas representaciones revelan sorprendentes niveles de conciencia acerca de los aspectos sociales, éticos y políticos de la actividad científica y de los métodos y prácticas de los científicos.

Con todo, lo que parece quedar claro es que, si bien esta imagen estereotipada se forma tempranamente, a medida que la escolaridad avanza, los rasgos más característicos se acentúan con fuerza (Dibarboure, 2010), con el correlato correspondiente del desinterés por las asignaturas científicas por parte de los jóvenes y el consiguiente estancamiento de la matrícula en las carreras científicas, en Iberoamérica en general y en Argentina en particular:

> Entre los años 2008 y 2010, y bajo la coordinación del Observatorio CTS de la OEI, se aplicó una encuesta representativa de la población estudiantil de nivel medio en algunas capitales, ciudades y sus ámbitos periféricos, de Iberoamérica. Un tercio de los estudiantes iberoamericanos que participaron en la encuesta dijo que le gustaría estudiar una carrera vinculada al área de las ciencias sociales. Un 16% mencionó carreras vinculadas a las ingenierías y tecnologías. Y sólo un 5% se inclinó por las ciencias exactas y naturales. Las ciencias en general tampoco gozan, por otra parte, de una aceptación amplia entre las nuevas generaciones al momento de imaginar el futuro profesional. En la misma encuesta con estudiantes iberoamericanos, solamente uno de cada diez dijo que el trabajo de los científicos podría ser de su interés. Preguntados sobre la motivación que podrían tener en una carrera científica sus pares generacionales, un tercio de los jóvenes dijo que la ciencia no era atractiva para los jóvenes. (Polino y Chiappe, 2011: 10-11)

Si bien la exigua matriculación en carreras científicas y tecnológicas es una problemática multicausal, no reducible a la desidentificación del estudiantado con la ciencia, no resulta nada trivial enfocarse en cómo una imagen estereotipada de la ciencia y de los científicos contribuye a inhibir la vocación científica:

"Estas visiones deformadas obturan la posibilidad de una alfabetización científica genuina, alejando a muchas personas de las ciencias naturales y mitificando estas disciplinas. [La inhibición consecuente] [e]s un fenómeno triple, en el que intervienen los maestros, que les transmiten a los chicos que eso no es para todos; los padres, que en general piensan que es una profesión poco valorada socialmente, mal remunerada, no muy feliz para las mujeres, y los propios jóvenes, que internalizan esos mandatos y terminan pensando yo no soy para esto, es muy complicado, a mí no me da". (Adúriz-Bravo, en Stekolschik, 2008).

Como señala Adúriz-Bravo, esta exclusión además tiene una importante componente de género, relegando a las mujeres a un plano muy secundario. En este sentido, Jones y colegas (2000) advierten que la escuela –a pesar de ser un ambiente que no deja de estar afectado por prejuicios de género– debería ser el lugar donde los estudiantes puedan encontrar los mejores valores y actitudes acerca de la ciencia. Señalan que las profesoras y profesores no pueden eludir la responsabilidad de presentar la ciencia como igualmente apropiada para chicos y chicas, pretender que también las chicas sean capaces de utilizar las herramientas científicas con facilidad, y lograr que tanto chicos como chicas se involucren reflexivamente en las actividades científicas. Además, concluyen que continuar con el *statu quo* sin transformar la cultura es condenar a las estudiantes a permanecer en los bordes de la ciencia.

En esa misma línea, Mary Wyer (2003) afirma que la persistencia y el no abandono de las carreras científicas por parte de las mujeres están en estrecha relación con las imágenes positivas de científicos, con las actitudes positivas respecto de la igualdad de género, y con las experiencias positivas en las aulas de ciencias. La amplia instalación de los estereotipos que de alguna manera prescriben qué grupos pueden acceder y tener éxito en el ámbito del conocimiento científico resulta en una desidentificación por parte de los grupos aludidos respecto de la ciencia, que ya dan por sentada su falta de habilidad para ello. Esto es particularmente notorio en el caso de las minorías étnicas y de las mujeres, como bien señala Steele (1997).

**Enfoque metodológico**

Como se ha señalado en el apartado anterior, el uso de las narraciones donde se hace referencia a la actividad científica puede resultar un instrumento valioso a la hora de caracterizar la imagen de ciencia y de científico subyacente en la persona que construye la historia. De hecho, así ha sido considerado en algunos estudios llevados a cabo con estudiantes de secundaria, generalmente acompañados por entrevistas a los autores (Reis y Galvão, 2006, 2007). Es esta línea de análisis de narrativas la que hemos elegido para abordar los cuentos producidos por los y las jóvenes en lo que a imágenes de ciencia y de científico se refiere. En nuestro caso, no contamos con la posibilidad de entrevistar a los autores de los relatos (como sí han hecho las investigaciones que hemos citado precedentemente), lo que hubiera permitido quizás profundizar y aclarar algunas visiones. Sin embargo, creemos que nuestra muestra cuenta con elementos a favor, en el sentido de que quienes escribieron los cuentos lo hicieron más que nada motivados por un interés genuino de participar con sus producciones, y no "conminados a hacerlo" para un estudio posterior, lo cual permitiría suponer en nuestro caso una mayor puesta en juego de sus propias visiones acerca de la ciencia y los científicos.

Nuestro abordaje ha sido eminentemente cualitativo, con la intención de rastrear en los textos recurrencias a las categorías que Manzoli y colaboradores (2006) refieren como dimensiones de análisis (figura 3):

--- Figura 3 más o menos aquí ---

*La dimensión social*: Interesa conocer en qué medida las narrativas muestran la actividad científica como resultado de la labor conjunta de diferentes personas e instituciones, con multiplicidad de intereses y valores, y al alcance de todo el mundo (géneros, clases sociales, culturas, edades), o bien la presentan como empresa individual, solitaria, aislada, principalmente masculina y reservada a unos pocos privilegiados.

*La dimensión práctico-tecnológica*: Comprende las alusiones a la práctica científica, si es que aparecen los elementos clásicos de la iconografía científica: tubos de ensayo, alambiques, fórmulas, jeringas, que se corresponden principalmente con la labor experimental, o bien dispositivos tecnológicos: computadoras, instrumentos de observación, máquinas, dispositivos variados inventados "ad hoc", o bien una actividad científica extramuros o intramuros a través de metodologías diversas no experimentales.

*La dimensión de conocimiento*: Cómo es que se presenta en las narraciones el conocimiento producido o puesto en acción (Castelfranchi, 2003):

1. el conocimiento como una violación: el "descubrimiento" de aquello que no debía ser descubierto o la incursión en cuestiones prohibidas;

2. el conocimiento como poder (para el bien o para el mal) y el peligro de la pérdida de control de la creación;

3. el conocimiento como control de la naturaleza y la transformación de lo inanimado en animado[2].

*La dimensión espacial-temporal*: Lo espacial alude por un lado a los ambientes donde se desarrolla la actividad científica, esto es, la imagen del científico encerrado en su laboratorio, su estudio, su habitación, o bien perteneciendo a ámbitos específicos de instituciones donde realiza su investigación (empresas, universidades, entes gubernamentales). Por otro, a la actividad extramuros que caracterizaría lo que damos en llamar el "científico de campo". En cuanto a lo temporal, esta dimensión hace referencia tanto al tiempo que invierte el científico hasta llegar a los resultados pretendidos, como también a la ubicación temporal de la narración: describiendo hipotéticos sucesos en el pasado histórico o ubicando la acción en especulativos escenarios futuros, estos últimos muchas veces vinculados con acontecimientos del presente.

*La dimensión ética*: Puede asumir características que vinculen a la actividad científica como promotora del bienestar humano o bien como la culpable de desgracias y peligros insoslayables. En este sentido, pueden ser reflejados en esa dicotomía no solo la actividad *per se*, sino los productos de la misma y, por sobre todas las cosas, los sujetos involucrados en ella. Aparece aquí también la puja entre el "deber ser" de la ciencia y lo que la ciencia realmente es.

---

[2] Estos tres aspectos, tan vinculados entre sí y tan relacionados con la dimensión ética, han sido bien descriptos, aludiendo a la metáfora del Gólem, por Harry Collins y Trevor Pinch (1996).

*La dimensión emotiva-mítica*: Bastante emparentada con las anteriores, en esta dimensión cobran importancia todos aquellos aspectos a los que se les confiere una perspectiva fuera de lo común: héroes o villanos, la trasgresión de los límites, los atributos cuasi mágicos, el "jugar a ser dioses"… todo ello acompañado por el premio o castigo correspondiente de acuerdo a lo actuado.

**Registros de evidencias en los textos a la luz de las dimensiones de análisis**

Una primera mirada al conjunto de cuentos permitiría distinguir diferentes abordajes de la ciencia por parte de sus autores. En muchas narraciones se alude a diversos tópicos científicos a partir de analogías. Así, por ejemplo, hay referencias a la Vía Láctea en términos de compararla con una ciudad y sus suburbios; a agujeros negros como habitaciones de las cuales nada sale; la antropomorfización de fenómenos naturales (espermatozoides en su carrera por alcanzar el óvulo, el canibalismo en las arañas). Otros tantos cuentos podrían encuadrarse en el género de ciencia-ficción, en sus términos más clásicos de proponer escenarios especulativos, relatando acontecimientos posibles, la mayoría de ellos con una temporalidad definida en el futuro o en el presente y algunos pocos en el pasado. Para nuestro análisis, utilizamos principalmente las narraciones correspondientes a este género, en tanto aparecen personajes que se dedican a la actividad científica, susceptibles de ser analizados en función de sus rasgos más característicos y frecuentes.

Del total de los 147 cuentos considerados, concretamente, 78 pueden ser encuadrados en esta última categoría. A partir del análisis argumental de los mismos, se evidencia que:
- 72 de ellos abordan la dimensión social,
- 63, la dimensión de conocimiento,
- 60, la dimensión ética,
- 58, la dimensión práctico-tecnológica,
- 52, la dimensión emotiva-mítica, y
- 38, la dimensión espacial-temporal.

*Acerca de la dimensión emotiva-mítica*

Hay en los cuentos una fuerte impronta de elementos mágicos, que remiten muy probablemente a la influencia de historias fantásticas, tanto en sus versiones literarias como en sus trasvases al cine, al estilo de Harry Potter, donde las luchas entre el bien y el mal se dirimen en base a poderes sobrenaturales. La ciencia aquí ocupa el papel de poder transmutador, que es patrimonio de unos pocos elegidos. Aparecen así referencias a pócimas, fórmulas ultrasecretas, invenciones de máquinas de increíble poder, posibilidades de producir cambios asombrosos:

> Llegó a mí como una revelación suprema, que desde un principio supe que me ayudaría a cambiar el mundo. Después de años de labor científica e investigación mi trabajo pudo ser concluido. (…) Mis esfuerzos dieron sus frutos, logré un buen día la satisfacción de encontrar la cura a todos los males. Descubrí que ciertos químicos y fluidos introducidos en el cuerpo daban a éste una vida perdurable. Yo fui el primero en usarlo y vi como efectivamente mi cuerpo se rejuvenecía.
>
> ("Los inmortales")

### *Acerca de la dimensión ética*

En estrecha relación con la dimensión emotiva-mítica se presenta la cuestión de jugar a transponer los límites permitidos acerca de qué es bueno y qué es malo, y cómo estos límites se desdibujan y se atraviesan cuando se pierde el control del poder que el conocimiento otorga.

En general, el propósito que guía obsesivamente al científico se relaciona con creaciones, invenciones o descubrimientos extraordinarios, destinados a cambiar el mundo de una vez y para siempre. Pero siempre está la amenaza latente de la peligrosidad. En muchos relatos, esta amenaza termina en concreción de la fatalidad. En muchos cuentos, esta puja entre el bien y el mal se encarna en diferentes grupos de personas con intereses diferentes:

> [N]uestro último hallazgo científico […] podía llegar a ser extremadamente peligroso. Habíamos creado un nuevo tipo de onda artificial […] que podía intervenir en el comportamiento de los seres vivos y permitía manejar su voluntad sin dejar rastros.
>
> ("Ondas peligrosas")

> –Jugaste a ser Dios, pero te salió mal, él es perfecto mientras que vos no.
>
> ("Hilos de sangre en el espejo")

–Todo esto fue obra de una mente maestra, que no apareció en ningún momento (…). El doctor simplemente fue un objeto.
–Pero, ¿cómo?
–Nanomáquinas. Son máquinas pequeñísimas que se dirigen desde una computadora… supongo que ésa. El doctor tenía una insertada en el cerebro. Parece que manipulan la actividad neuronal, teniendo un efecto alucinógeno que no permite al portador darse cuenta de lo que está haciendo. Al mismo tiempo, envían señales eléctricas, ya programadas, permitiendo a quien las maneja dirigir las acciones del portador.

<div style="text-align: right">("Nanocrimen")</div>

El hombre deseaba más, sumergido en el insaciable deseo de poder, digno de su especie: siempre pretendiendo controlarlo todo. La mujer, por su lado, se lamentaba en silencio, consciente de que el cariño que sentía por él la arrastraría por siempre contra la corriente de la razón.

<div style="text-align: right">("Dimensiones")</div>

–Verá, esta tecnología fue codiciada por muchos grupos de todo tipo, políticos, militares, etcétera, etcétera, etcétera. Y por ende, el mundo se ve en una terrible guerra.

<div style="text-align: right">("La academia de música de la calle Rodríguez")</div>

–[Al] PRG (…) puede definírselo en el cual se modifica la información del genoma humano para lograr una base de igualdad de derechos, eliminando por ejemplo, con la manipulación genética terapéutica todo riesgo de enfermedades genéticas antaño incurables. También, gracias a la manipulación eugenésica nuestros progenitores tienen el derecho natural y merecido de mejorar nuestros caracteres somáticos (…).
–¿Pero no cree que cada quien debe tener la libertad de decidir sobre su descendencia?
–(…) el no someterse al Proceso de Reorganización Genética implica un destino azaroso. Las personas naturales son en realidad personas vulnerables e incapaces de adaptarse a nuestra sociedad, segura y organizada. Incluso sus genes natos presentan una fuerte amenaza, si llegaran a mezclarse con alguno de los nuestros, podrían surgir catástrofes impensadas.

<div style="text-align: right">("La quimera y el sapo")</div>

> (…) la ciencia tiene su lado bueno y su lado malo. (…) Me corrijo (…) [l]a ciencia no es "buena o mala" sino que nosotros la disfrutamos o manipulamos…
>
> ("¿Y el título?... mejor me callo")

### *Acerca de la dimensión espacial-temporal y de la dimensión social*

Si bien aparecen en algunos cuentos los equipos de trabajo integrados por varias personas, la mayoría de las veces se trata de trabajo individual, solitario, aislado del resto del mundo, generalmente en un laboratorio o en otros espacios cerrados. Solamente el científico conoce los detalles de su investigación, que adquiere carácter secreto, críptico, enigmático. Su mente brillante, privilegiada, lo hace el único capaz de hacer lo que hace. Se dedica obsesivamente, a tiempo completo, a lograr su propósito. La mayoría de las personas que se describen realizando actividad científica son hombres.

La cuestión de la temporalidad en los cuentos se refleja en prefiguraciones que hablan de un futuro con dificultades, donde los problemas son diferentes de los del presente, pero originados en el hoy: enfermedades, graves desequilibrios ambientales, escenarios posibles amenazantes producto de los desarrollos sin control de la ciencia actual:

> Inicié un proyecto de investigación muy complejo, todo era producción mía y no me permitía la ayuda de nadie… [E]ra capaz de aislarme para realizar todas las cosas…
>
> ("Mas allá")

> –Tío, haré historia–, recuerdo que me dijo. (…) Al día siguiente, él se sirvió una taza de leche y se fue temprano para la reducida habitación de experimentos de mi padre. Luego, regresó por su colchón y se lo llevó a aquel lugar (…).
> –Tío, de aquí no saldré jamás.– (…) Méderic se encerró en ese ambiente. (…) Según su petición, lo mantuve a pan y agua por años. De acuerdo a lo que leí luego en sus anotaciones, había confeccionado innumerables fórmulas para la mejor comprensión de su entorno…
>
> ("Inmensidad")

### *Acerca de la dimensión de conocimiento y la práctico-tecnológica*

Las referencias a los modos de producción de conocimiento científico siempre se relacionan con un esfuerzo metódico, sostenido, incansable, que se centra en la observación y la experimentación. Se hace mención en muchos casos a hipótesis, teorías y leyes en el marco de las investigaciones, así como también aparecen modelos científicos con los que los jóvenes autores se han encontrado a lo largo de la escolaridad. Los atributos del lenguaje científico se expresan con frecuencia en términos de fórmulas y ecuaciones. Otro componente frecuente es presentar al científico como inventor, como creador de máquinas y dispositivos que nunca se sabe bien en qué basan su funcionamiento, pero que producen efectos asombrosos:

> Sólo el descubrimiento de la naturaleza del Universo tiene un sentido duradero. Ésa era la verdad para él, y a pesar de que era conciente de su superioridad intelectual, sabía que nunca entendería a los que pensaran lo contrario. Las agujas del reloj continuaban su recorrido y, pese a eso, el hombre seguía sentado, realizando anotaciones abstractas para cualquier persona, aunque no para él.
>
> ("El sentido duradero de la naturaleza del Universo")

**Una imagen de ciencia y de científico multidimensional**

Cabría preguntarse cuál es el origen de esta particular imagen de ciencia en la que se suele coincidir. Y una respuesta plausible puede llegar a encontrarse si acudimos a vincularla con la imagen del *alquimista*. Si bien este personaje ya es reconocible en la Edad Antigua y en diversas partes del mundo, es en el alquimista del Medioevo europeo donde se van a marcar aquellos aspectos herméticos y mágicos que lo harán trascender en los estereotipos. La alquimia se prefiguraba en ese entonces como la promesa de fabulosas riquezas, poder y longevidad: la piedra filosofal permitiría a quien la hallara transmutar metales en oro, y el elixir de la juventud acabaría con la amenaza de la muerte (Eliade, 1983).

Los alquimistas persistieron hasta muy avanzado el Renacimiento, aún hasta la Modernidad (Vickers, 1984). Los practicantes de la alquimia solían trabajar ocultos en lugares secretos, como los sótanos de los castillos. La mayoría de sus actividades se realizaban de noche: muchos de sus procedimientos requerían de la luz de la Luna. Moviéndose siempre en la zona gris entre lo natural y lo sobrenatural, utilizaban un lenguaje oscuro, esotérico, plagado de símbolos, como forma de preservar sus conocimientos. Paracelso, uno de los principales alquimistas del siglo XVI, que puso a la alquimia al servicio de la medicina, decía de sus pares: "Se entregan diligentemente a su labor", "no pierden el tiempo en conversaciones ociosas, sino que encuentran su felicidad en el laboratorio". Les aconsejaba:

> Aprended pues la Alquimia, también llamada Spagyria, y ella os enseñará a discernir lo falso de lo verdadero. Con ella poseeréis la luz de la Naturaleza y con ella por tanto podréis probar todas las cosas claramente, discurriéndolas de acuerdo a la lógica y no por la fantasía, de la que nada bueno puede resultar.[3]

La literatura de ficción del siglo XIX se ha encargado de retomar las particularidades de estos protocientíficos. El *Fausto* de Goethe y el *Frankenstein* de Mary Shelley son ejemplos de cómo se conjugan las fuerzas naturales y las fuerzas místicas en pos de una obsesión: ser como dioses, jugar a la inmortalidad, a la creación de vida. Resulta especialmente sugerente la explicación que propone Joachim Schummer (2006) acerca de por qué (o para qué) se muestra en la literatura del siglo XIX esta imagen. Según este autor, se crea como una respuesta literaria a la emergencia de la nueva química, que se erige como el prototipo de las ciencias experimentales vistas como una seria amenaza a la unicidad del conocimiento y relacionadas con el ateísmo, el materialismo, el nihilismo y con su pretendida arrogancia. Para resaltar los aspectos negativos y atacar las ideas iluministas de la ciencia, los escritores se basan en la figura del *alquimista medieval*. Esta imagen del personaje encerrado, que invierte su propia vida para conseguir aquello que se propone, dejando de lado todo lo mundano, incluso su propia familia, o bien la del truhán que, prometiendo enseñar estas técnicas, esquilma a sus víctimas, despojándolas de sus bienes.

En el *Fausto* de Goethe se reconoce la potencialidad de la experimentación química con todo su poder creador, pero también su falta de capacidad para comprender y evaluar sus artefactos, de lo que deviene la pérdida de control de las creaciones y, por tanto, su peligrosidad casi ineludible. En *La búsqueda de lo absoluto*, Balzac pone en primer plano la arrogancia del científico al tratar de emular el poder divino a partir de la química, lo que lo lleva a la ruina y a perder su familia. En *Frankenstein, o el moderno Prometeo*, Shelley personifica en el Dr. Viktor Frankenstein al científico amante de la alquimia, admirador de

---
[3] Paracelso, *Opus paramirum*, Libro Uno: Causas y orígenes de las primeras sustancias, Capítulo Tercero: Sobre el modo de acción de las primeras substancias, el sujeto intermedio y la alquimia. En: *Obras completas*, pág. 102.

Paracelso, encarnando una química de finales del siglo XVIII, una química moderna como la nueva portadora de la piedra filosofal.

Así,

> [p]ara hacer que la arrogancia fuera una acusación moralmente convincente para sus lectores, los autores del siglo XIX crearon al "científico loco". Transformado a partir del alquimista ya establecido en la literatura medieval, el científico loco combina la arrogancia con toda la perversión moral que los escritores del siglo XIX podían imaginar. (…) [E]sta figura literaria ha dominado la visión pública de la ciencia desde entonces. (Schummer, 2006: 125; la traducción es nuestra)

**El rescate del estereotipo desde otras manifestaciones culturales**

Ahora bien, si partimos del presupuesto de que es poco probable que gran parte de los jóvenes autores/as de los cuentos que nos ocupan hayan leído estas obras literarias del siglo XIX, ¿cómo se explica que reflejen en sus cuentos una imagen de científico con tantos puntos de contacto con ese *científico loco*? Desde ya que el cine en principio plasma en la pantalla muchas de esas obras de la literatura de ficción del siglo XIX (*Fausto*, *Frankenstein*, "El extraño caso del Dr. Jekyll y Mr. Hyde", entre otras), y después la impronta de este "científico loco" pasa a formar parte de una mirada ya clásica del cine sobre las personas que se dedican a la ciencia, con diferentes matices ("El profesor chiflado", "Volver al futuro", "El hombre invisible", etc.) (Moreno Lupiáñez, 2003; Weingart et al., 2003; Guerra Retamosa, 2004).

También dan cuenta de este estereotipo los cómics (Gallego, 2007), los dibujos animados, la TV, y la publicidad (Campanario et al., 2001; Medina Cambrón et al., 2007). Asumimos que este rescate que hacen los *mass media* de la figura estereotípica la hace más cercana al estudiantado. Si bien por un lado la proliferación de referencias a la ciencia y la actividad científicas en los medios va en constante aumento, muchas veces dicha abundancia informativa carece de rigurosidad, apelando al sensacionalismo, exagerando hallazgos, o soslayando el carácter provisional del conocimiento científico y el lugar para la incertidumbre, anunciando supuestas verdades reveladas, y mostrando a los científicos estereotipadamente, como magos o héroes (Schäffer, 2011).

En lo que respecta a la televisión y los dibujos animados, es indudable que la gran cantidad de tiempo que pasan los niños, niñas, adolescentes y jóvenes frente a las pantallas deja su impronta, para bien o para mal, según sea el caso. Por ejemplo, numerosas investigaciones reseñan que generalmente se presentan pocos programas donde se muestren científicas, siendo el panorama general dominado por la figura del científico varón, dotado de especial inteligencia. Cuando aparecen mujeres, lo hacen en un rol secundario o su opinión es subestimada, dando a entender que poseen menos capacidades y habilidades para la actividad científica (Long y Steinke, 1996; Weingart et al., 2003; Vilchez-González y Palacios, 2006; Steinke et al., 2008). De hecho, estudios recientes señalan el impacto positivo en la identificación de las adolescentes con la ciencia a partir de programas que muestran tanto a científicos como a científicas con similares desempeños. No obstante, dichos programas siguen sugiriendo que la actividad científica está reservada a personas brillantes, dedicadas a tiempo entero a su tarea, y casi sin vida familiar y social (Long et al., 2010).

**La imagen de científico y la ciencia que se enseña y que se aprende en las aulas**

Nuestra hipótesis de trabajo es que la imagen de científico constituye un *epifenómeno* de una imagen de ciencia subyacente. Es decir, los estudiantes –y el público en general– "corporizan" en estos personajes su propia imagen de ciencia (Adúriz-Bravo, 2005a).

En tren de caracterizar esa imagen de ciencia subyacente, se podría coincidir en que surge de una visión marcadamente empiroinductivista, que considera la ciencia como construcción ahistórica, marcadamente individualista, independiente de valores, ideologías, intereses y contextos, y por tanto neutral, objetiva, sin dudas, infalible y dueña de la verdad. Al mismo tiempo se muestra como una empresa elitista y exclusora, esencialmente masculina, fundada en una racionalidad científica centrada en un único método. Suele acentuarse su carácter críptico y hermético, que sólo puede ser descifrado por verdaderos "iniciados".

La investigación internacional suele catalogar estas imágenes de ciencia como deformadas, distorsionadas o inadecuadas (Chen et al., 1997; Hodson, 1998; Adúriz-Bravo, 2001; Manassero Mas y Vázquez Alonso, 2001; Fernández et al., 2002; Hugo y Adúriz-Bravo, 2003; Vázquez et al., 2006; Demirbaş, 2009). En muchos de los trabajos a los que hicimos referencia se afirma que estas visiones deformadas se transmiten cuando se enseña ciencia (por ejemplo, en Fernández et al., 2002). Como se señalara anteriormente, existe evidencia de que los rasgos estereotípicos acerca de la ciencia y los científicos se acentúan con el decurso de la escolaridad (Dibarboure, 2010). Algunas investigaciones recientes (Pujalte et al., 2011a) parecerían indicar que las niñas y los niños del nivel preescolar e inicial poseen representaciones acerca de la ciencia mucho más ricas y variadas, en las que es muy potente la autoidentificación con la figura del científico, en el marco de diversos escenarios posibles para la actividad científica. Paulatinamente, a medida que transcurre la escuela primaria y luego la secundaria, esas representaciones acaban pareciéndose mucho a la imagen de ciencia y de científico del profesorado. Esto revela la importancia de conocer la imagen de ciencia que traen las profesoras y los profesores de ciencias, en formación y en servicio, para poder intervenir sobre ella (Pujalte et al., 2011b).

**A modo de conclusión**

Estudiamos en detalle una serie de cuentos con temática científica producidos por jóvenes estudiantes de los últimos años de la escuela secundaria. El primer objetivo de este trabajo fue indagar cuál es la representación acerca de la ciencia que surge de las narraciones a partir de cómo caracterizan particularmente a las personas que se dedican a la actividad científica. En ese sentido, podríamos sostener que los cuentos de ficción producidos por los jóvenes de nuestra muestra revelan unas imágenes de ciencia y de científico alejadas del saber metacientífico actual, lo que es coincidente con lo se viene reflejando en las investigaciones realizadas en esta línea.

Una segunda meta propuesta consistió en establecer los nexos que permitirían vincular dichas representaciones con otras manifestaciones culturales. En ese sentido, nos permitimos sugerir la impronta de los medios masivos de comunicación, con su insistencia en rasgos estereotípicos fuertemente pregnantes, y cómo estos medios se nutrirían a partir de la literatura de ficción decimonónica y su rescate de la figura del alquimista medieval.

Finalmente, tratamos de echar luz sobre la importancia que tiene conocer cuáles son las imágenes de ciencia y de científico del estudiantado en relación con la procura de una educación científica de calidad para todas y todos. Podríamos afirmar que estas imágenes se constituyen en un obstáculo para alcanzar esa meta, entre otras cosas porque se corresponden con un desinterés por las asignaturas científicas y la merma consiguiente en la matrícula en las carreras de ciencia y tecnología. En este último sentido, volvemos a resaltar (como muchos otros autores) la importancia que tiene conocer la imagen de ciencia que sustenta el profesorado, en tanto se trata a nuestro criterio de una de las principales fuentes de donde se nutre la visión de ciencia de los jóvenes, representados en este caso por los autores y autoras de los cuentos.

A partir de este estado de situación, habría que procurar el diseño de intervenciones didácticas que promuevan imágenes más inclusivas de la actividad científica. Para ello, el foco debería ser puesto en la *enseñanza* de la naturaleza de la ciencia, en los diferentes niveles educativos, con un énfasis especial en la formación inicial y continuada del profesorado de ciencias (Adúriz-Bravo, 2005b).

**Referencias**

Figura 1

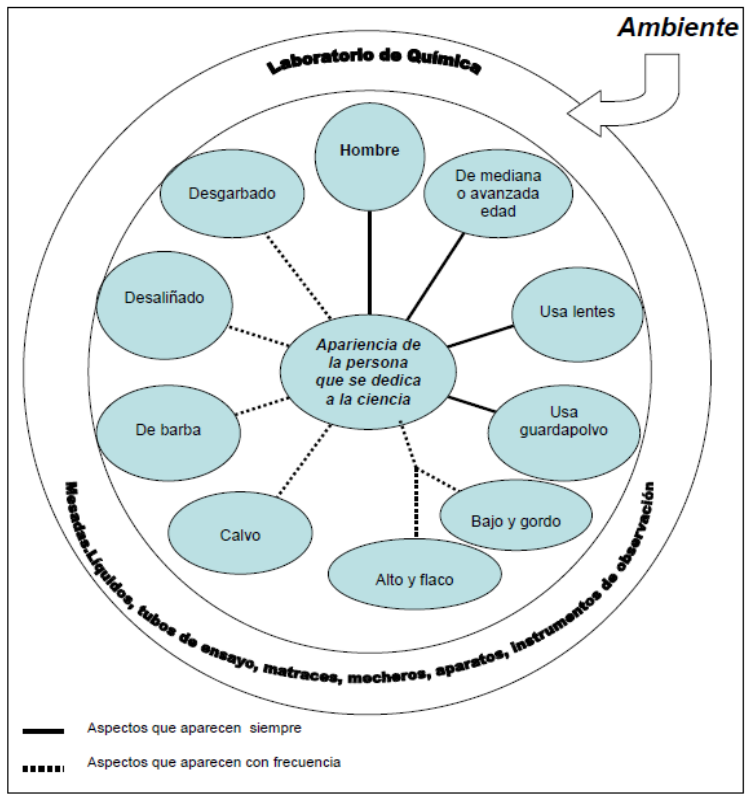

**Figura 1.** Aspectos recurrentes respecto de la apariencia de las personas que se dedican a la actividad científica y de las características del ambiente donde suelen trabajar, según la indagación de Mead y Metraux (1957).

Figura 2

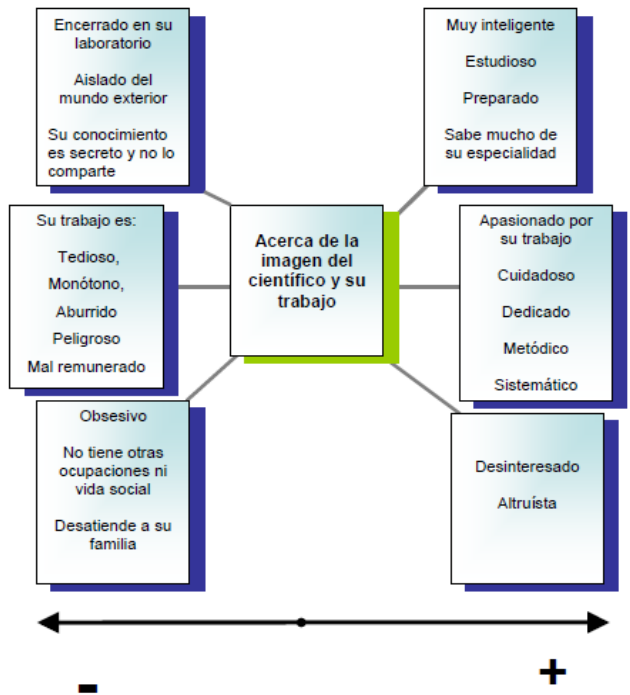

**Figura 2.** Aspectos positivos y negativos de la imagen de las personas que se dedican a la ciencia, según la indagación de Mead y Metraux (1957).

Figura 3

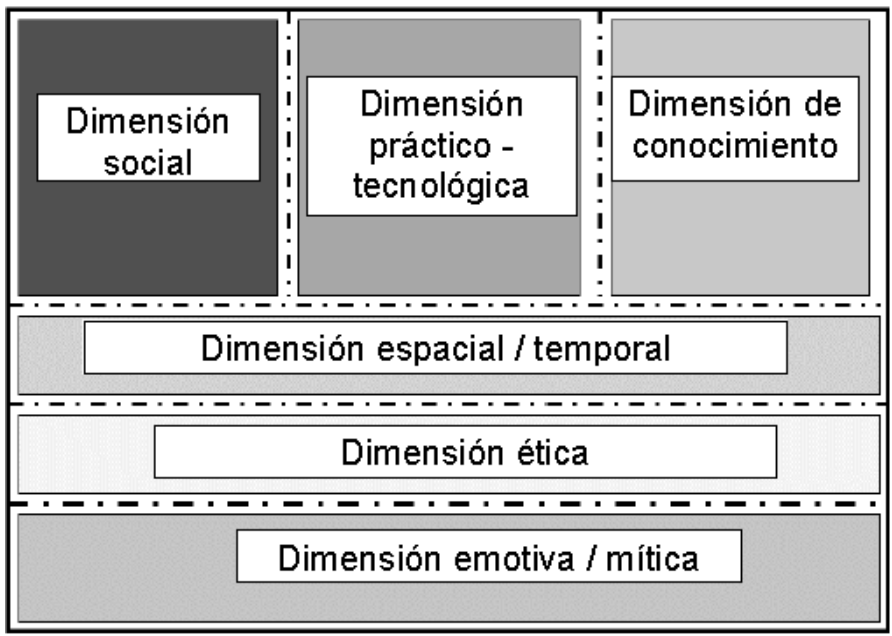

**Figura 3.** Dimensiones de análisis (Manzoli et al., 2006).